\documentclass [preprint] {revtex4}

\usepackage[dvips]{graphicx}
\usepackage{amssymb}

\def\beq {\begin{equation}}
\def\eeq {\end{equation}}
\def\bea {\begin{eqnarray}}
\def\eea {\end{eqnarray}}
\def\ni {\noindent}
\def\nn {\nonumber}

\def\lp {\left( }
\def\rp {\right) }
\def\lb {\left[ }
\def\rb {\right] }
\def\lc {\left\{ }
\def\rc {\right\} }

\def\rar {\rightarrow}

\def\Db {\bar{\Delta}}

\def\Lb {\bar{L}}

\def\Ob {\bar{\Omega}}

\def\ct {\tilde{c}}

\def\sp {\!+\!}
\def\sm {\!-\!}

\def\cK {{\cal{K}}}

\def\a {\alpha }
\def\b {\beta}
\def\d {\delta}
\def\D {\Delta}
\def\e {\epsilon}

\def\g {\gamma}
\def\l {\lambda }

\def\o {\omega}
\def\p {\pi}

\def\bq {\mbox{\boldmath $q$}}
\def\bQ {\mbox{\boldmath $Q$}}

\def\bD {\mbox{\boldmath $\Delta$}}

\begin{document}

\title{Unitarization and low-energy scattering data}

\author{ P. C. Magalh\~{a}es}
\email[]{patricia@if.usp.br}
\author{M. R. Robilotta}
\affiliation{ Instituto de F\'{\i}sica, Universidade de S\~{a}o Paulo,  
S\~{a}o Paulo, SP, Brazil}
 
\date{\today }

\begin{abstract}

A procedure based on the well known $\cK$-matrix formalism
is presented, which makes 
patterns in inelastic regions of low-energy scattering data
considerably more transparent. It relies on the use of an empirical kernel, obtained by eliminating elastic loops from the experimental amplitude. This allows structures associated with resonances,
such as locations, widths and heights,  to become visible
with the naked eye.
The method is illustrated with a study of the
$P$-wave $K\p$ amplitude.
\end{abstract}

\pacs{...}

\maketitle
\section{introduction}
The problem of understanding the dynamical content of mesonic
scattering amplitudes, with energies ranging from threshold 
to a few GeV, is interesting by itself. 
Nevertheless, it became especially urgent in recent times, 
because these amplitudes play important roles 
in final state interactions  which occur after
primary weak decays of quarks within $D$ and $B$ mesons.
Decays of heavy mesons are presently object of intense scrutiny and 
the understanding of patterns observed in Dalitz plots require
accurate descriptions of both primary weak vertices and 
hadronic final state interactions\cite{final,BR}.

In non-relativistic quantum mechanics, the standard treatment
of elastic scattering begins with a potential, which is fed
into the Schr\"odinger equation.
In the elastic regime, the potential is real and 
the dynamical equation yields complex wave-functions, that can be
parametrized in terms of phase shifts. 
When inelasticities are present, one needs a complex optical 
potential and the corresponding damped wave-functions
can be parametrized by their phase shifts and moduli or inelasticity parameters. 
The formulation of the scattering problem for relativistic
particles is roughly similar.
Dynamics is now described by the Bethe-Salpeter  equation
and its {\em kernel} corresponds to the potential.
In the case of scattering of two generic particles $X$ and $Y$,
this kernel amounts to the sum of all {\em proper} intermediate diagrams,
that is, those which cannot be split into two pieces by cutting
$X$ and $Y$ propagators only. 
Similarly to what happens with the potential,
real kernels give rise to elastic interactions,
whereas inelasticities require complex kernels.

In quantum mechanics, the so called inverse problem, in which 
one attempts to derive potentials from scattering amplitudes, 
has been widely studied and one knows that it does not 
have unique solutions. 
The same holds for the relationship between scattering amplitudes
and kernels, in the framework of relativity.
This problem has been discussed long ago by Lang\cite{Lang},
who has shown that constraints imposed by unitarity
on scattering amplitudes cannot fix kernels completely.   
However, in the case of low and medium energy meson-meson scattering, 
the Bethe-Salpeter equation can be simplified and there is 
a window in which 
an approximate solution to the inverse problem does exist,
which we discuss in the sequence.
In the last two decades, this simplification of the Bethe-Salpeter 
has been extensively used in the dynamical unitarization of amplitudes
and successfully applied to a wide number of 
problems\cite{UNIT, Khemchandani}.

\begin{figure}[h]
\begin{center} 
\includegraphics[width=0.35\columnwidth,angle=0]{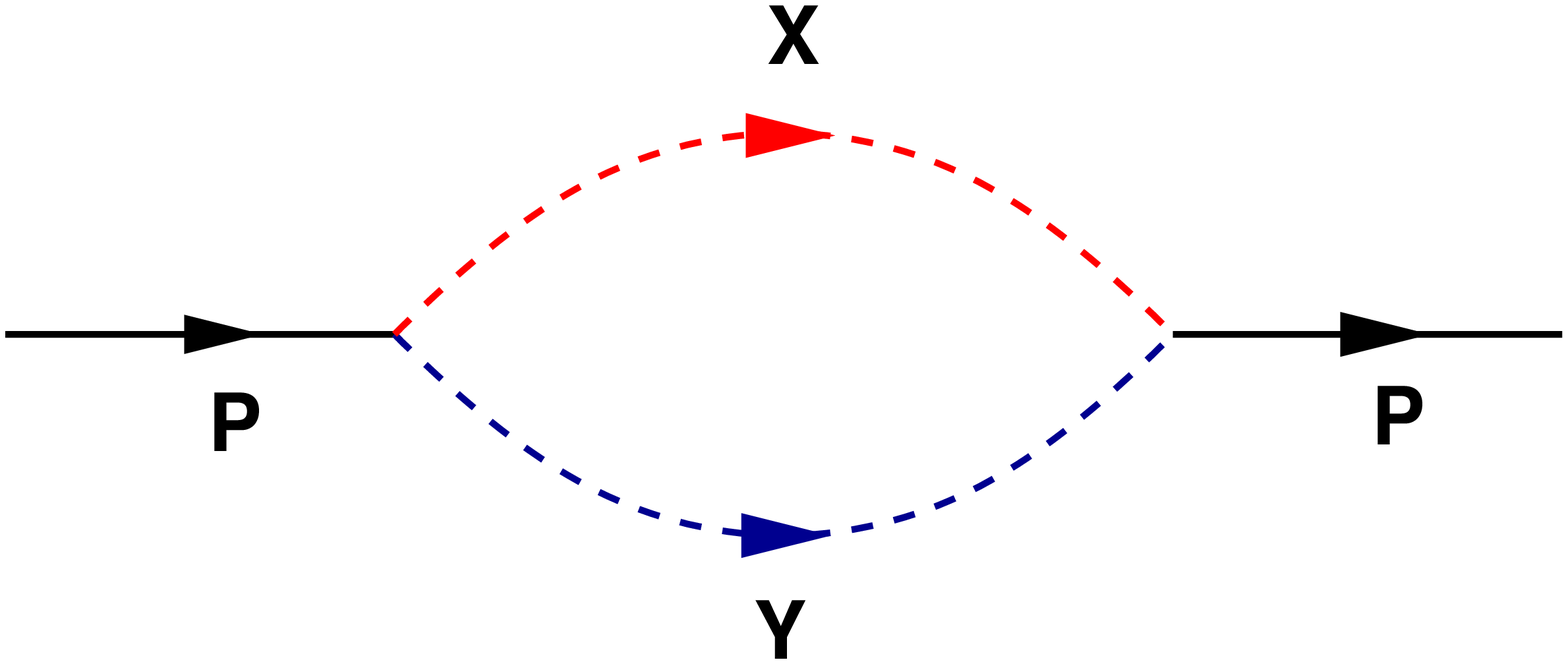}
\end{center}
\caption{Light two-particle propagator.}
\label{F-light}
\end{figure}

Quite generally, the reaction 
$X(p_x)\,Y(p_y) \rar X(p'_x) \, Y(p'_y)$ 
is described in terms of Mandelstam variables $s, t, u$ and,
in the center of mass system (CM), results can be expressed in terms 
of the three-momentum $\bq$, with
$\bq^2 = s\, \rho^2/4$,
\bea
\rho &\!=\!& \sqrt{1 - 2\,(m_x^2 \sp m_y^2)/s + (m_x^2 \sm m_y^2)^2/s^2}\,.
\label{m1}
\eea
The corresponding amplitude $T_{xy}$ can be formally represented 
in terms of the Bethe-Salpeter equation as
\bea
T_{xy}(s)&=& \cK_{xy}(s) 
- i \int \frac{d^4 \ell}{(2\p)^4}\; 
\frac{\cK_{xy}(P,\ell) \;\; T_{xy}(P,\ell)}
{[(\ell \sp P/2)^2 \sm m_x^2 \sp i\,\e\,] \;
[(\ell \sm P/2)^2 \sm m_y^2 \sp i\,\e\,]} \;,
\label{m2}
\eea
where $K_{xy}$ is the kernel and $P=p_x \sp p_y\,$.
This equation can be simplified if one assumes that, at  low energies, the dependences 
of both $\cK_{xy}(P,\ell)$ and $T_{xy}(P,\ell)$ on the loop 
variable $\ell$ can be neglected.
This assumption has been used in Ref.\cite{Lang},
tested in many instances\cite{UNIT, pelaezOO},
and allows the amplitude to be written as
\bea
T_{xy}(s) = \cK_{xy}(s)\;\lb 1 - T_{xy}(s)\; \Omega_{xy}(s) \rb \;,
\label{m3}
\eea
where the function
\bea
\Omega_{xy}(s) = i\,\int \frac{d^4\ell}{(2\p)^4}\;
\frac{1}{[(\ell \sp P/2)^2-m_x^2\,]\;[(\ell \sm P/2)^2-m_y^2\,]}
\label{m4}
\eea
represents the elastic two-particle $XY$ 
propagator, shown in fig.\ref{F-light}.

\begin{figure}[h]
\hspace{-20mm}
\includegraphics[width=.5\columnwidth,angle=0]{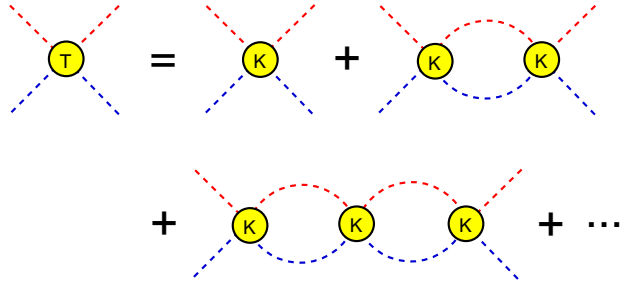}
\caption{Amplitude for the process $XY \rar XY$ represented
as a geometric series involving the kernel $\cK$
and light-meson propagators.}
\label{F-iter}
\end{figure}

The integral (\ref{m4}) is ultraviolet divergent and a subtraction
is needed.
Following ref.\cite{GL}, we write the regular part of $\Omega_{xy}$ as
\bea
&& \Ob_{xy}(s)= \Omega_{xy}(s)-\Omega_{xy}(0)\;,
\label{m5}
\eea
where the divergent part is contained in $ \Omega_{xy}(0)$.
Regularization amounts to replacing this term 
by an unknown finite constant $C$, which
has to be fixed later.
The function $\Ob$ can be evaluated analytically and,
above threshold, one has
\bea
\Ob_{xy}(s) &\!=\!& (\Ob_{xy})_R + i\,(\Ob_{xy})_I
=-  \Lb_{xy}(s)/ (S_{xy}\,16 \p^2 )\;,
\label{m6} \\[2mm]
\Lb_{xy} &\!=\!& 1 
+ \frac{m_x^2 + m_y^2}{m_x^2 - m_y^2}\;\ln \lb\frac{m_x}{m_y}\rb
- \frac{m_x^2-m_y^2}{s}\;\ln \lb \frac{m_x}{m_y} \rb  
\nn\\[2mm]
&\!-\!& \,\frac{\sqrt{\l}}{s}\;
\ln \lb \frac{s- m_x^2- m_y^2 + \sqrt{\l}}
{2\,m_x\,m_y}\rb 
+ i\,\p\;\frac{\sqrt{\l}}{s} \;,
\label{m7}
\eea
with $\sqrt{\l} = s\, \rho$ and $S_{xy}$, the symmetry factor, 
is 1  for $X\neq Y$ and 2 for $X=Y$.
After regularization, eq.(\ref{m3}) may be solved for $T_{xy}$
and one finds\cite{Lang} 
\beq
T_{xy}(s) = \frac{\cK_{xy}(s)}{1 + [\Ob_{xy}(s) \sp C]\; \cK_{xy}(s)} \;.
\label{m8}
\eeq
In the elastic regime, the kernel $\cK_{xy}$ is real and the 
complex structure of eq.(\ref{m7}) makes $T_{xy}$ 
automatically unitary.
On the other hand, when inelasticities are present, the kernel $\cK_{xy}$ 
acquires imaginary components.
The conceptual relationship between the amplitude and the kernel 
can be better understood by rewriting eq.(\ref{m8}) as the 
geometric series
\bea
T_{xy}(s) &\!=\!& \cK_{xy}(s)
- \cK_{xy}(s) \,[\Ob_{xy}(s) \sp C]\, \cK_{xy}(s)
\nn\\
&\!+\!& \cK_{xy}(s) \, [\Ob_{xy}(s) \sp C] \,\cK_{xy}(s) \,
[\Ob_{xy}(s) \sp C] \, \cK_{xy}(s) + \cdots  \;.
\label{m9}
\eea
This shows that the amplitude is just a monotonous iteration of the
kernel, as shown in fig.\ref{F-iter}.
It is therefore desirable to know the kernel, 
since it encompasses all relevant dynamical effects.

Data are usually presented by means of a non-relativistic amplitude $f$,
which can be parametrized in terms of either phase and modulus or
phase and elasticity, 
and related with the relativistic amplitude $T$ by 
\bea
\frac{\rho\; T_{xy}}{16 \p}
= f_{xy} = |\,f_{xy}\,|\;e^{i\,\d_{xy}} 
=\frac{1}{2i} \lb \eta_{xy} e^{2i \bar{\d}_{xy}}-1\rb \;.
\label{m10}
\eea
Feeding the experimental amplitude $T_{xy}^{exp}$, extracted from
eq.(\ref{m10}), into eq.(\ref{m8}), and inverting it,
one finds a solution to the inverse problem, given by
\beq
\cK_{xy}^{exp}(s) 
= \frac{T_{xy}^{exp}(s)}{1 - [\Ob_{xy}(s) \sp C]\; T_{xy}^{exp}(s)} \;.
\label{m11}
\eeq
As we discuss in the sequence, this function $\cK_{xy}^{exp}$ 
conveys dynamical information 
in a very transparent way, since it is free from $XY$ 
intermediate states.
It rests at the basis of the procedure proposed here.

In order to evaluate eq.(\ref{m11}), one needs
to fix the constant $C$. 
Since it cannot be fixed directly from data, it must be 
taken as a free parameter and the function $\cK_{xy}^{exp}$ is 
contaminated by this freedom. 
However, once a choice is made, the kernel is uniquely determined and
its use in eq.(\ref{m8}) yields results which do not depend
on conventions.
The choice made in Ref.\cite{Lang} corresponds to $C=0$. 
Here, we fix $C$ by noting that, at the the point $s=s_{\p/2}$, 
for which $\d_{xy}(s_{\p/2})=\p/2\,$, 
one has $T_{xy}(s_{\p/2})= i\, 16\p/\rho(s_{\p/2})$.
Using this, together with the result $(\Ob_{xy})_I=-\rho/16\p$, one finds
\beq
\cK_{xy}^{exp}(s_{\p/2}) = -\,\frac{1}{[\Ob_{xy}(s_{\p/2})]_R \sp C} \;.
\label{m12}
\eeq
One then chooses $C=-[\Ob_{xy}(s_{\p/2})]_R\,$, because this gives rise 
to a pole in  $\cK_{xy}^{exp}(s_{\p/2})$, correlated with the phase 
$\d_{xy}(s_{\p/2})=\p/2\,$ and associated with a light 
resonance $R_L$ in the $XY$ channel, of mass $m_{R_L}=\sqrt{s_{\p/2}}$.

\begin{figure}[h]
\hspace{-40mm}
\includegraphics[width=.4\columnwidth,angle=0]{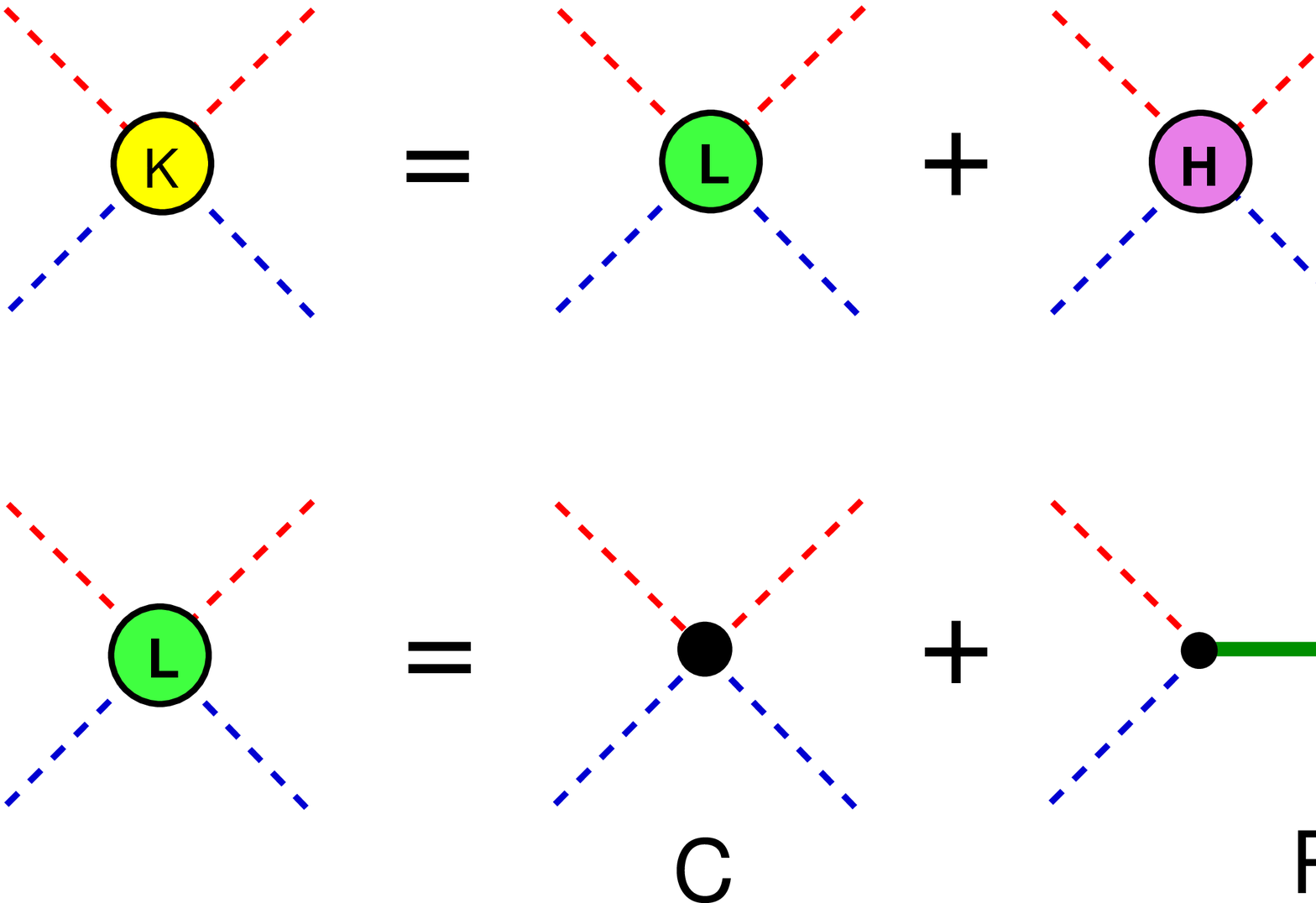}
\caption{Top: decomposition of the kernel $\cK$ into light and heavy 
components $\cK_L$ and $\cK_H$.
\\
Bottom: chiral model for $\cK_L$, based on a leading contact 
interaction (C), a bare $s$-channel light resonance $(R_L)$
and correction terms $(t+u)$.}
\label{F-LH}
\end{figure}

As far as theory is concerned, it is useful to decompose the kernel 
$\cK$ into light and heavy pieces, denoted by $\cK_L$ and $\cK_H$,
as in Fig.\ref{F-LH}.
The latter includes interactions of heavy systems and 
is discussed in App.\ref{heavyres}. 
The light component $\cK_L$ is real and involves 
only light-particles.
In the framework of chiral models\cite{EGPR}, this light kernel
is given by the sum of a leading contact term,
a light bare $s$-channel resonance $R_L$, and 
corrections associated with exchanges of resonances
in $t$- and $u$-channels.
We stress, however, that the use of a chiral model  is not compulsory.
The only essential ingredient in $K_L$ is the light resonance,
whereas the other terms could well be replaced by a
polynomial in $s$, with free coefficients.
In order to stress this feature it is convenient to express the
light kernel as
\bea
\cK_L = \cK_{R_L} + \cK_{{BG}_L} \;.
\label{m12x}
\eea
Hence the theoretical kernel reads
\bea
\cK_{xy}^{th} \equiv \cK_{R_L} + \cK_{BG_L} + \cK_H \;,
\label{m18}
\eea 
The form of the bare resonance term $\cK_{R_L}$ is given 
in App.\ref{vecamp} and a chiral model for $\cK_{{BG}_L}$,
suited to the $K\p$ system, is presented in App.\ref{BG}.

\section{$K\p$ amplitude}
\label{kapi}

The preceding discussion, which is somewhat abstract, 
can be made more concrete with the help of a guiding instance.
With this purpose in mind, we present the case of $K\p$ scattering, 
for which data is available\cite{LASS} in the range 
$0.825 < E < 1.960$ GeV, and concentrate 
on the $P$-wave channel with isospin $1/2$.  
Data for this channel indicate that $\sqrt{s_{\p/2}}=0.894\,$GeV,
which corresponds to the mass of the $K^*(892)$.

\begin{figure}[h]
\hspace*{-45mm}
\includegraphics[width=0.5\columnwidth,angle=0]{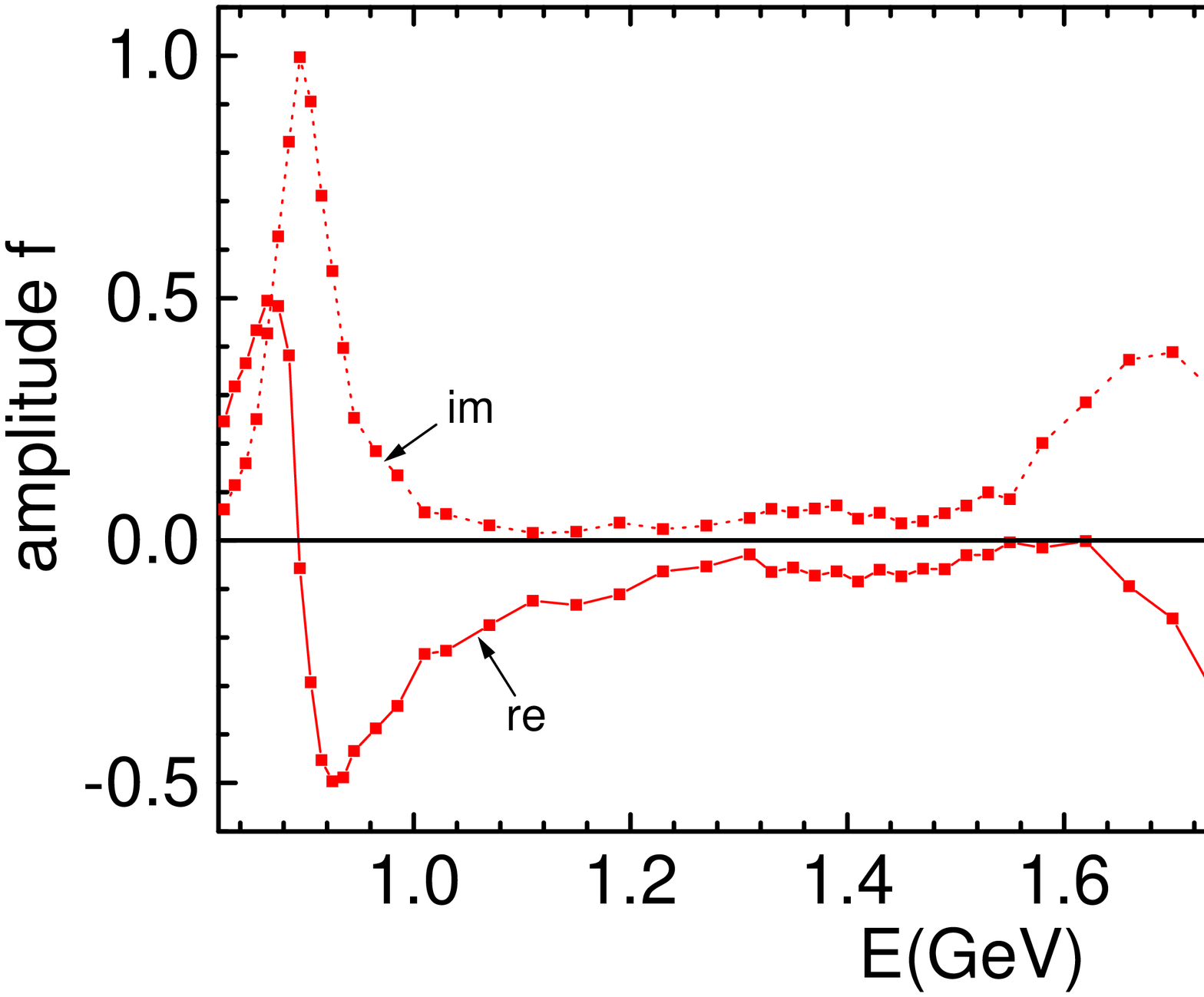}
\caption{Real and imaginary components of experimental
function $f_{K\p}$, eq.(\ref{m10});
lines were added to guide the eye.}
\label{F-fexp}
\vspace{5mm}
\hspace*{-45mm}
\includegraphics[width=0.5\columnwidth,angle=0]{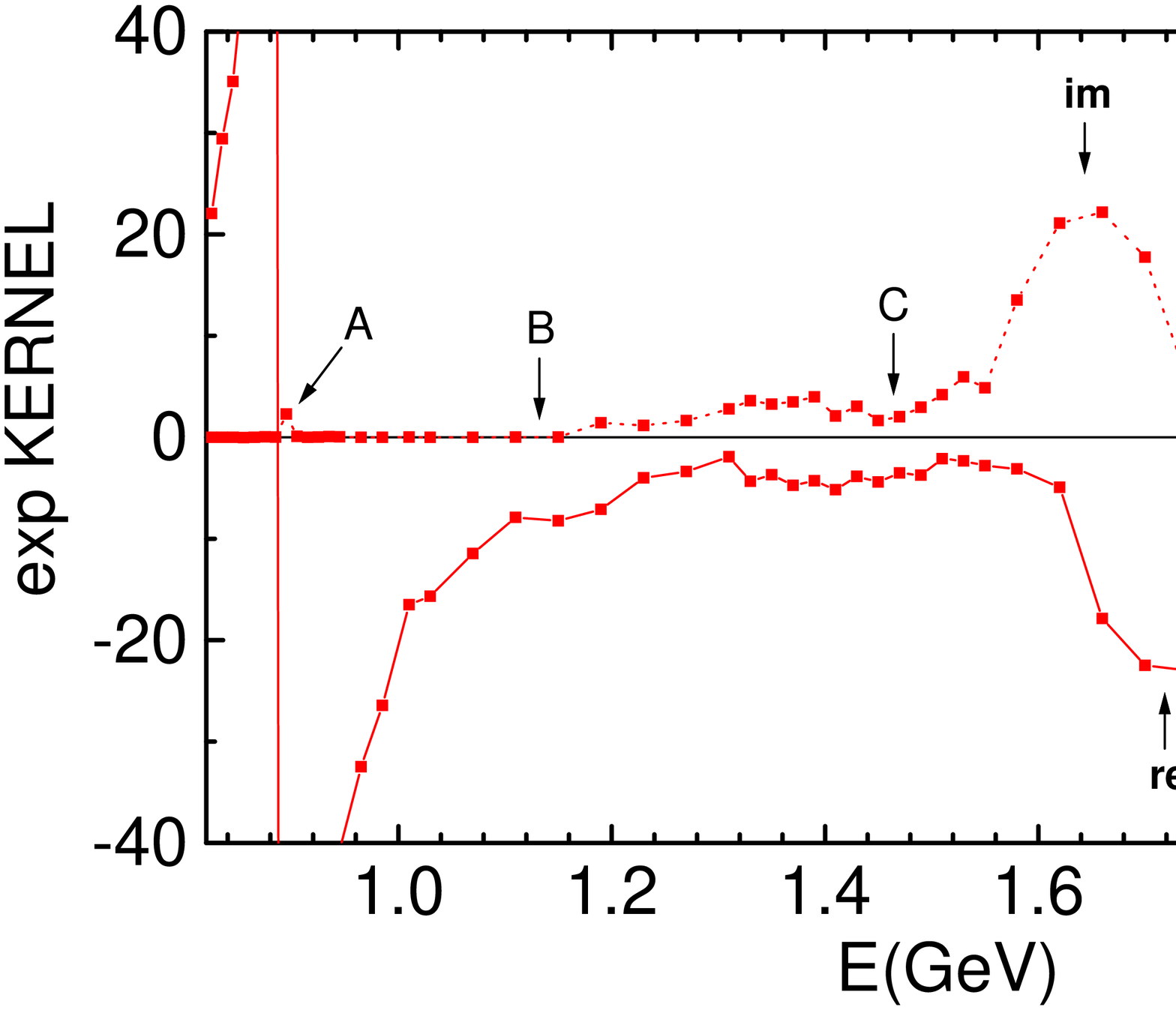}
\caption{Real and imaginary components of the function $\cK_{K\p}^{exp}$,
eq.(\ref{m11});
lines were added to guide the eye.}
\label{F-Kexp}
\end{figure}

\begin{figure}[h]
\hspace*{-45mm}
\includegraphics[width=0.5\columnwidth,angle=0]{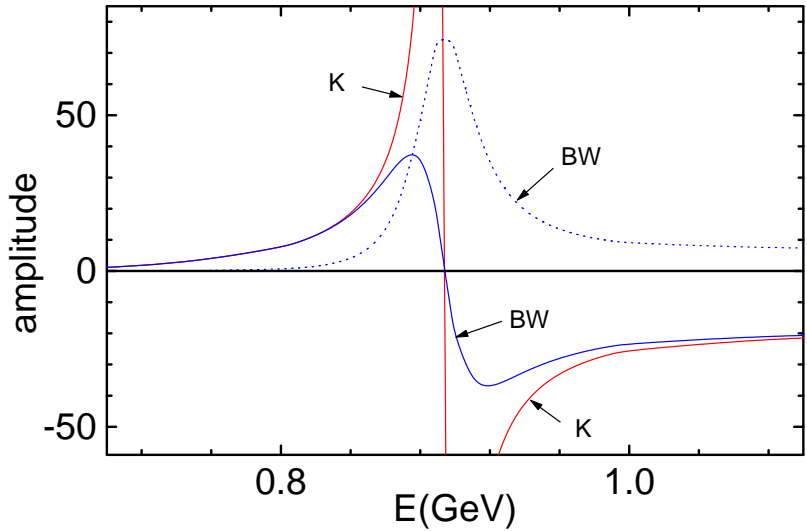}
\caption{Kernel (K, red curve) and real (BW, blue continuous curve)
and imaginary (BW, blue dotted curve) components
of the Breit-Wigner amplitude associated with the $K^*(892)$.}
\label{F-BWpole}
\end{figure}

In fig.\ref{F-fexp} we show the central values 
for real and imaginary components of the 
amplitude $T_{K\p}^{exp}$ and, in fig.\ref{F-Kexp},
the counterparts for $\cK_{K\p}^{exp}$,
obtained by using these values into eq.(\ref{m11}).
Comparing both figures in the region around $m_{K^*}$,
one notes a behaviour similar to a Breit-Wigner function (BWF)
for the amplitude and a bare pole for the kernel.
This is to be expected, since eq.(\ref{m8}) does indeed 
transform bare poles into Breit-Wigner functions and,
in this process, a resonance in the elastic regime acquires its width,
through  coupling to $K$ and $\p$ states. 
The kernel associated with $m_{K^*}$ and the corresponding BWF are shown in fig.\ref{F-BWpole}.
In the case of $P$-waves, the kernel describing 
the propagation of a single vector resonance is presented in 
appendix \ref{vecamp}.
In the framework of the $K\p$ system, eq.(\ref{a5}) reads
\bea
&& \cK_{R_L} = \cK_{K^*} =  - \,\lc g_{K^*} \, |\bq|\,\rc^2 \;
\frac{s}{s \sm m_{K^*}^2} \;,
\label{m13}
\eea
where $g_{K^*}=G_V/F^2$ 
is the $K^*K\p$ coupling constant\cite{EGPR}.
When a structure of this kind is used into eq.(\ref{m8}), one gets
a typical Breit-Wigner amplitude.
The width of this resonance is directly proportional to $g_{K^*}^2$,
stressing its dynamic origin. 

This  pattern is fully backed by field theory.
Effective lagrangians, suited to describing low energy mesonic
interactions, have been intensively studied in the last three 
decades\cite{GL, Wchi} and the treatment of resonances has been 
thoroughly discussed in ref.\cite{EGPR}.
Within  the lagrangian formalism, the only intrinsic property of a 
resonance is its mass, which is a real parameter.
The width is not intrinsic and has to be generated dynamically, 
by means of couplings to possible decay products.

The elimination of $K\p$ intermediate states from data yields the 
kernel $\cK_{K\p}^{exp}$, shown in fig.\ref{F-Kexp}, where the $K^*(892)$ pole in the real
component is clearly visible, at point A.
Inspecting the imaginary component, one learns that it vanishes 
up to point B, indicating the higher end of the elastic regime.
From that region onwards, reactions such as
$K\,\p \rar K^*\,\p $ or $K\,\p \rar K\,\rho$ become possible\cite{PDG12}.
At higher energies, one finds at least two bumps, associated with 
resonances.
One might also be tempted to interpret data beyond point D as the tail
of a third bump. 
As loops are the only source of imaginary components in hadronic field 
theory, an  imaginary part in  $\cK_{K\p}^{exp}$ is a clear indication
of the presence of loops involving heavy intermediate states.

The extraction of information from  bumps in the imaginary component 
requires some theory, since one has to understand how basic dynamics
allows several resonances to get dressed.
A somewhat detailed discussion of this subject
can be found in appendix \ref{heavyres}.
Results given there are consistent with unitarity indicating that, 
as expected,
the imaginary part of $\cK_{K\p}^{exp}$ cannot be negative.
Inspecting the region close to point D in fig.\ref{F-Kexp},
one learns that a problem with data may be present there.
A comprehensive fit function for $\cK_H$, the heavy component of the kernel, is given in eq.(\ref{b8}).
However, this function is complicate and we consider two simpler limits,
whereby either each resonance develops its own imaginary part independently 
or this is a fully collective process.
The former case corresponds to the direct sum of two Breit-Wigner
functions, given by eq.(\ref{b9}), and the latter, to eq.(\ref{b10}),
in which bare propagators are added before the unitarization
procedure.
The distinctive feature of the kernel given by eq.(\ref{b10}) is that 
it has to vanish at a point at the right of the first peak,
as shown in fig.\ref{F-BWD}.
This could, in principle, be useful for describing the region
around point D in fig.\ref{F-Kexp}.

In our $K\p$ instance, the structure of the function $\cK_H$ has 
been chosen as a combination of eqs.(\ref{b9}) and (\ref{b10}), 
which reads
\bea
\cK_H = |\bq|^2 
\lc \l_a^2 \, \frac{\D_a}{1- i\,g_a^2\,\D_a \,|\bQ_1|^3/8\p\sqrt{s} }
+ \xi^2 \, 
\frac{g_{b}^2 \,\D_b + g_{c}^2\, \D_c}
{1 - i\,[g_b^2\,\D_b + g_c^2\,\D_c]\,|\bQ_2|^3/8\p\sqrt{s}} \rc \;,
\label{m15}
\eea
where $\l_a\,$, $\xi\,$, and the $g_i$ are free parameters
and 
\bea
\D_i = -\, \frac{s}{s - m_i^2} \;.
\label{m16}
\eea
The denominators in eq.(\ref{m15}) are complex and the same happens
with the entire function $\cK_H$.
This choice was motivated by fig.\ref{F-Kexp}, since it seems 
to describe a broad resonance at energies around $1.4\,$GeV,
associated with $\D_a$, and a dip  around $1.8\,$GeV,
associated with the weighted sum of $\D_b$ and $\D_c$.

A possible model for the real background function $\cK_{{BG}_L}$ 
is given in appendix \ref{BG}. 
It receives contributions from a contact interaction $\cK_L^C$,
associated with chiral symmetry\cite{GL},
and exchanges of scalar and vector resonances in 
$t$ and $u$ channels, denoted by $\cK_L^S$ and $\cK_L^V$.  
Our trial function for the light part of the kernel 
is written as
\bea
\cK_L =
\cK_{K^*} + \cK_{{BG}_L} = \cK_L^C +
\a \; \lb - \frac{G_V^2}{F^4} \, 
\frac{s\,|\bq|^2}{s \sm m_{K^*}^2}  + \cK_{K^*}^V \rb  
+ \b \; \cK_{L\rho}^V + \g \; \cK_L^S \;,
\label{m17}
\eea
where  $\a$, $\b$ and $\g$ are free parameters.

\begin{figure}[h]
\hspace*{-25mm}
\includegraphics[width=0.7\columnwidth,angle=0]{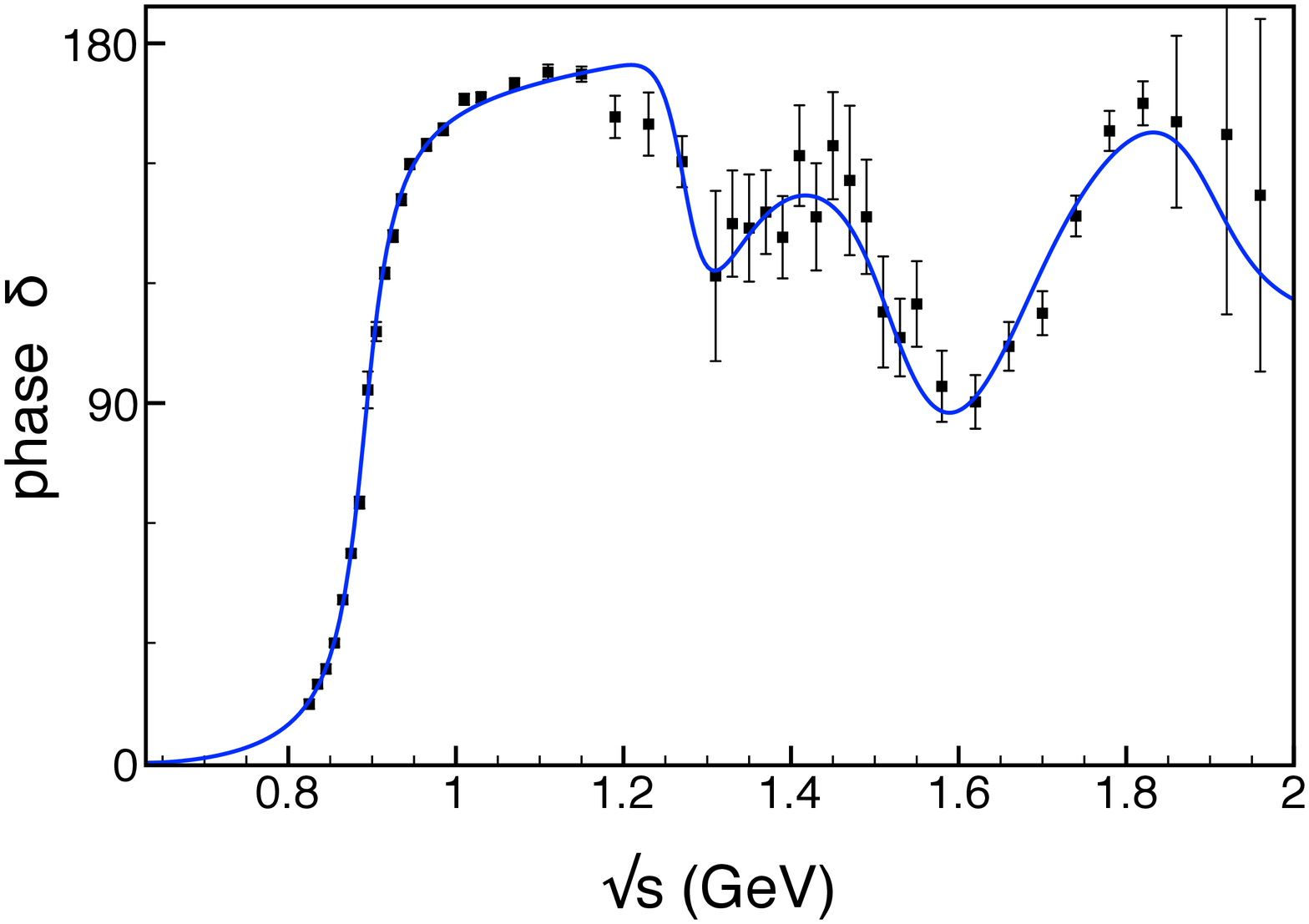}
\\[5mm]
\hspace*{-25mm}
\includegraphics[width=0.7\columnwidth,angle=0]{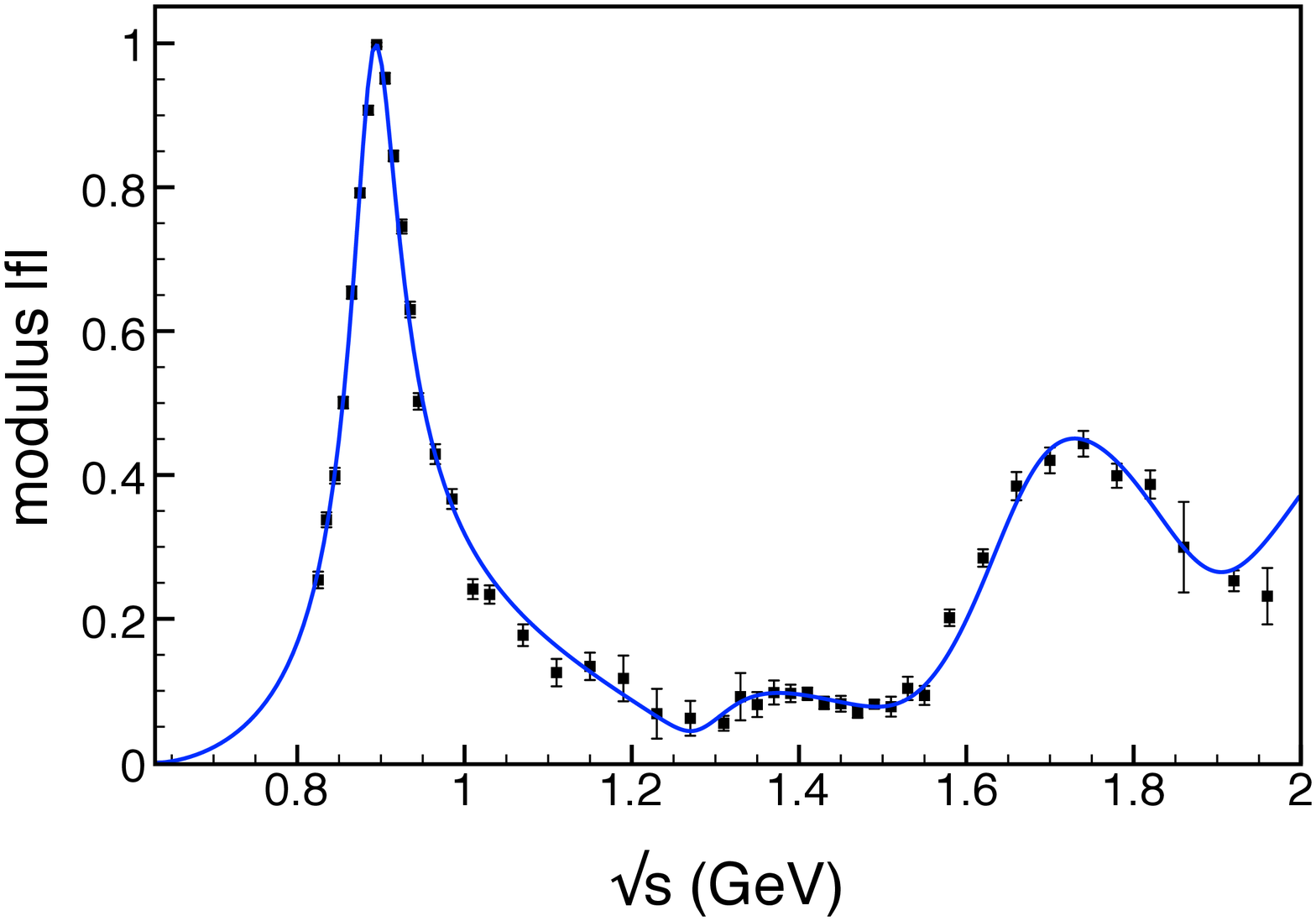}
\caption{Results for the  phase $\d$ (top) and modulus $|f| $(bottom)
of the  fitted $K\pi$  amplitude, 
using eqs (\ref{m15}) and (\ref{m17}), compared with data\cite{LASS}. The   $\chi^2/n.d.f.$ of the fit is  $2.13$ for the phase and $5.16$ for the modulus.}
\label{F-fitPH}
\end{figure}

Using the structures given by eqs.(\ref{m15}) and (\ref{m17}),
we fitted the phase and modulus for  the $P$-wave $K\p$ amplitude with isospin $1/2$ and the results are  given in fig.\ref{F-fitPH}.  We also obtained  the couple  of functions  phase-inelasticity  as given in fig.\ref{F-fitMO}.  The fit information and  the values for free parameters can be found in appendix \ref{values}.
In spite of the fact that data may have problems around 
point D of fig.(\ref{F-Kexp}), these results can be considered
as being satisfactory.


\begin{figure}[h]
\hspace*{-25mm}
\includegraphics[width=0.7\columnwidth,angle=0]{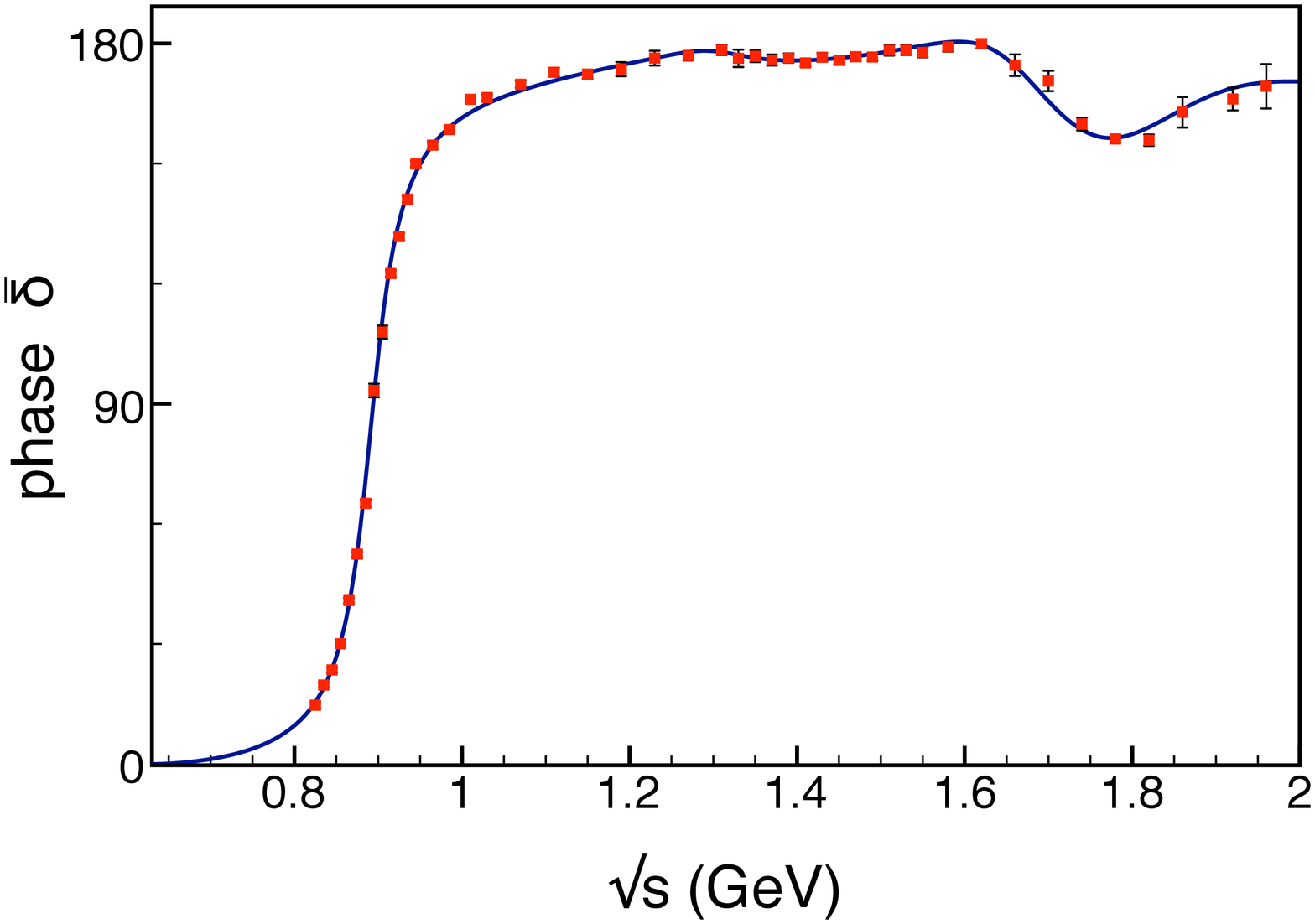}
\\[5mm]
\hspace*{-25mm}
\includegraphics[width=0.7\columnwidth,angle=0]{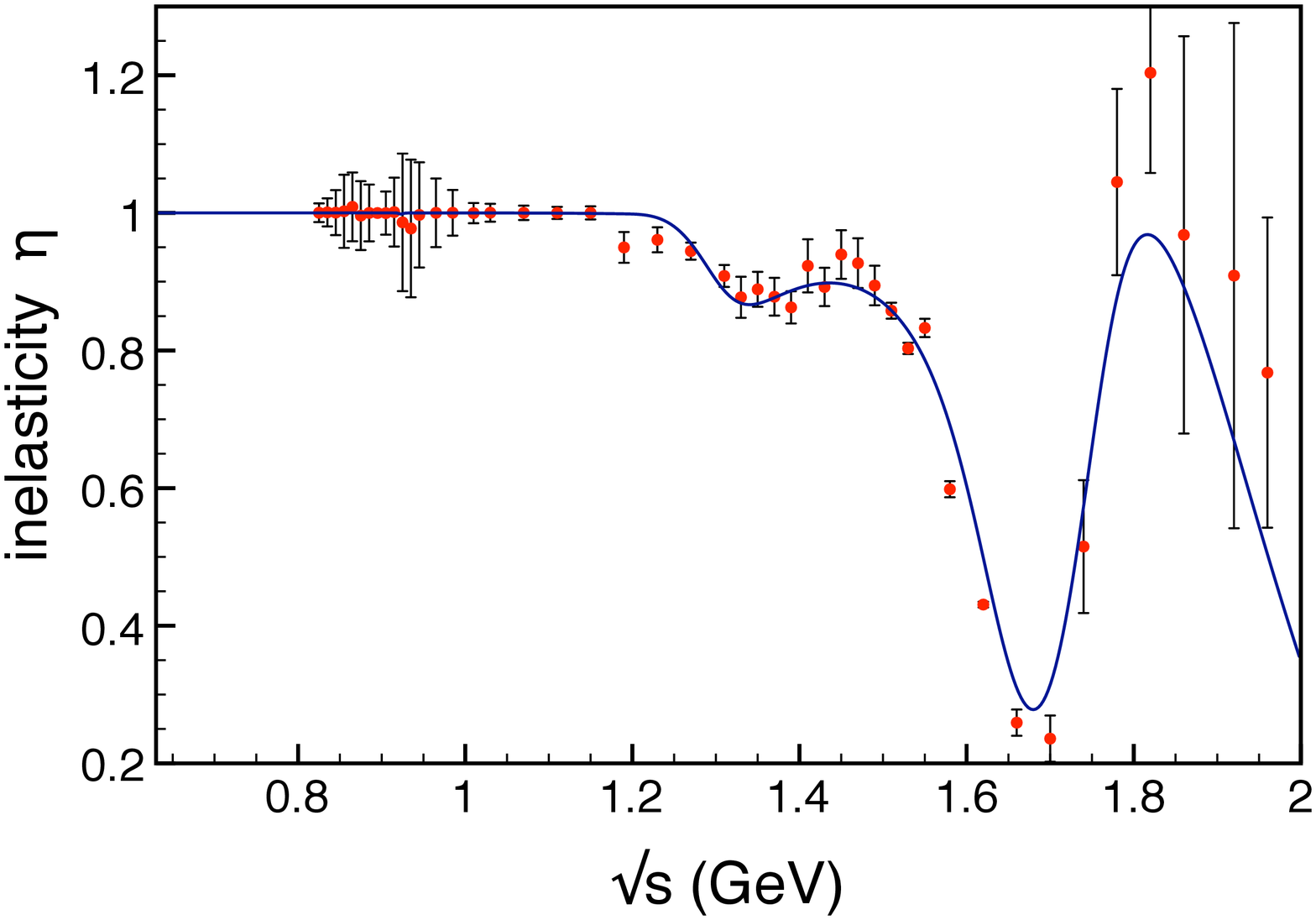}
\caption{Results for the  phase $\bar{\d}$ 
and inelasticity $\eta$ 
of the   fitted $K\pi$ amplitude, 
using eqs (\ref{m15}) and (\ref{m17}).}
\label{F-fitMO}
\end{figure}

\section{summary}
\label{fit}

We described a procedure which may be instrumental in
analyses of low and medium energy hadronic scattering data.
The basic idea is that $\cK$, the kernel of the system, 
which conveys dynamical information,
has a much simpler and transparent structure than the 
amplitude $T$.
In particular, Breit-Wigner trial functions are much more
suited to describing the relatively small complex structures
associated with possible inelasticities  rather
than the full interplay between real and imaginary 
components of the amplitude. 

The basic steps of this procedure are summarized below.
In the generic process $XY \rar XY$, one departs
from the scattering amplitude $T^{exp}_{xy}$ taken from data and, using
the $XY$ propagator $\Ob_{xy}$, given by 
eqs.(\ref{m6}) and (\ref{m7}),
into eq.(\ref{m11}), derives a representation 
for the experimental kernel $\cK_{xy}^{exp}$.
This function, in turn, can be decomposed as in eq.(\ref{m18}), which reads 
\bea
\cK_{xy}^{th} \equiv \cK_{R_L} + \cK_{BG_L} + \cK_H \;,
\nn
\eea 
where $\cK_{R_L}$ is real, contains the pole of the lowest resonance, 
and can be fixed by the neighbourhood of the point at which the phase 
reaches $90^0$.
The term $\cK_H$ is complex, conveys information about heavier
resonances, and can be represented as a combination of Breit-Wigner
functions suited to the case.   
Finally, $\cK_{BG_L}$ is a real contribution which may be either
fitted by 
a polynomial or associated with a model.
Once the kernel $\cK_{xy}^{th}$ is known, 
the full complex structure 
of the amplitude is recovered by using eq.(\ref{m8}).

\section*{ACKNOWLEDGMENTS}
The work of P.C.M. was supported by Funda\c{c}\~{a}o de 
Amparo \`{a} Pesquisa do Estado de S\~{a}o Paulo (FAPESP). 

\newpage
\appendix
\section{vector amplitude}
\label{vecamp}

\begin{figure}[h]
\hspace*{-25mm}
\includegraphics[width=0.3\columnwidth,angle=0]{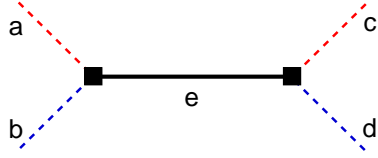}
\caption{Kernel involving an intermediate vector resonance.}
\label{F-BWDy}
\end{figure}

In the framework of $SU(3)$, the $s$-channel tree amplitude $T_{ab\rar cd}$, 
describing the reaction among pseudoscalars 
$P_a(p_a)\,P_b(p_b) \rar P_c(p_c)\,P_d(p_d)$, mediated
by a vector resonance $V_e$ of mass $m_e$ 
is given by\cite{EGPR} 
\bea
T_{ab \rar cd} &\!=\!& - f_{abe}\,f_{cde} \, G_{abe}\,G_{cde} \;
\frac{1}{s \sm m_e^2}\;
\nn\\[2mm]
&\! \times \!&
\lb 2\,s\,t + s^2 - s\,(M_a^2 \sp M_b^2 \sp M_c^2 \sp M_d^2)
+ (M_a^2 \sm M_b^2)\,(M_c^2 \sm M_d^2) \rb \;,
\label{a1}
\eea
where the $f_{ijk}$ are $SU(3)$ structure constants and the $G_{ijk}$
are coupling constants.
In the center of mass frame, this result can be rewritten 
in terms of initial and final tree-momenta $\bq$ and $\bQ$ as
\bea
T_{ab \rar cd} &\!=\!& - f_{abe}\,f_{cde} \, G_{abe}\,G_{cde} \;
\frac{4\, s}{s \sm m_e^2}\; \bq \cdot \bQ \;,
\label{a2}\\[2mm]
\bq &\! = \!& \frac{1}{2 \sqrt{s}}
\sqrt{s^2 - 2s\,(M_a^2 \sp M_b^2) + (M_a^2 \sm M_b^2)^2} \;,
\label{a3}\\[3mm]
\bQ &\! = \!& \frac{1}{2 \sqrt{s}}
\sqrt{s^2 - 2s\,(M_c^2 \sp M_d^2) + (M_c^2 \sm M_d^2)^2} \;.
\label{a4}
\eea
Projecting out the $P$-wave, one finds the typical structure 
for the kernel, which is given by 
\bea
\cK_{ab \rar cd} &\!=\!& - \,  
\lc g_{abe}\,|\bq|\,\rc \; \lc g_{cde} \,|\bQ|\,\rc \;
\frac{s}{s \sm m_e^2} \;,
\label{a5}
\eea
where a factor $4/3$ and the structure constants $f_{ijk}$ have been 
absorbed into the new coupling constants $g_{ijk}$.

In the case $X\!=\!a\!=\!c$, $Y\!=\!b\!=\!d$, this result
represents the $s$-channel pole of the bare light resonance $R_L$,
shown in Fig.\ref{F-LH}.
Its contribution to the light kernel $\cK_L$ is 
\bea
&& \cK_{R_L} =  \lc g_{R_L} \, |\bq|\,\rc^2 \; \D_{R_L} \;,
\label{a6}\\[2mm]
&& \D_{R_L} =  -\; \frac{s}{s \sm m_{R_L}^2} \;,
\label{a7}
\eea
where $g_{R_L}$ is the $XYR_L$ coupling constant and $\bq$ is given by eq.(\ref{a3}).
The iteration of this kernel, as in Fig.\ref{F-iter}, gives rise to a geometric series,
whose sum is indicated by $\bar{\cK}_{R_L}$ and reads
\bea
\bar{\cK}_{R_L} &\!=\!&  \lc g_{R_L} \, |\bq|\,\rc \lc \Db_{R_L} \rc 
\lc g_{R_L} \, |\bq|\,\rc \;,
\label{a8}\\[2mm]
\Db_{R_L} &\!=\!& \lc \D_{R_L} - \D_{R_L} [U_{xy}] \D_{R_L} 
+ \D_{R_L} [U_{xy}] \D_{R_L} [U_{xy}] \D_{R_L} + \cdots \,\rc  \;,
\label{a9}\\[2mm]
U_{R_L} &\!=\!& \lc g_{R_L} \, |\bq|\,\rc \,(\Ob_{xy} \sp C)\, 
\lc g_{R_L} \, |\bq|\,\rc \;,
\label{a10}
\eea
where $\Db_{R_L}$ is the dressed resonance propagator.
Summing the series (\ref{a9}) and using eq.(\ref{a7}),
one may express it as
\bea
\Db_{R_L} &\!=\!& \frac{\D_{R_L}}{1+ U_{R_L}\,\D_{R_L}} 
= -\,\frac{s}{s-m_{R_L}^2 - U_{R_L}\,s} \;,
\label{a11}
\eea
indicating that the resonance has been dressed.
The real part of $U_{R_L}$ influences its 
mass and the imaginary part gives rise to its width.

\section{heavy resonances}
\label{heavyres}

The amplitude describing the generic elastic reaction $XY \rar XY$
is given by the geometric series displayed in Fig.\ref{F-iter}
and involves both light and heavy kernels,
as indicated in Fig.\ref{F-LH}.
The light loop, shown in fig.\ref{F-light}, involves 
just particles $X$ and $Y$ themselves and is
described by the function $\Ob_{xy}$ given in eq.(\ref{m6}).

In this Appendix we describe the structure of the heavy kernel $\cK_H$, 
which encompasses all heavy loops, that is,
those different from the light one.
In order to make our discussion more concrete,
we assume that heavy loops represent the propagation 
of excited states, such as $X Y^*$ and $X^* Y$.
This assumption simplifies the presentation,
but results do not depend on it and remain valid 
if other intermediate states are included.

In $\cK_H$, the incoming and outgoing light  
particles $X$ and $Y$ couple to heavy systems, which are 
assumed to be {\em proper}, in the sense that they cannot 
be split into two parts by cutting intermediate $X$ and $Y$ lines only.
These heavy systems contain heavy loops, that give rise to 
complex components into $\cK_H$. 

\begin{figure}[h]
\hspace*{-10mm}
\includegraphics[width=.9\columnwidth,angle=0]{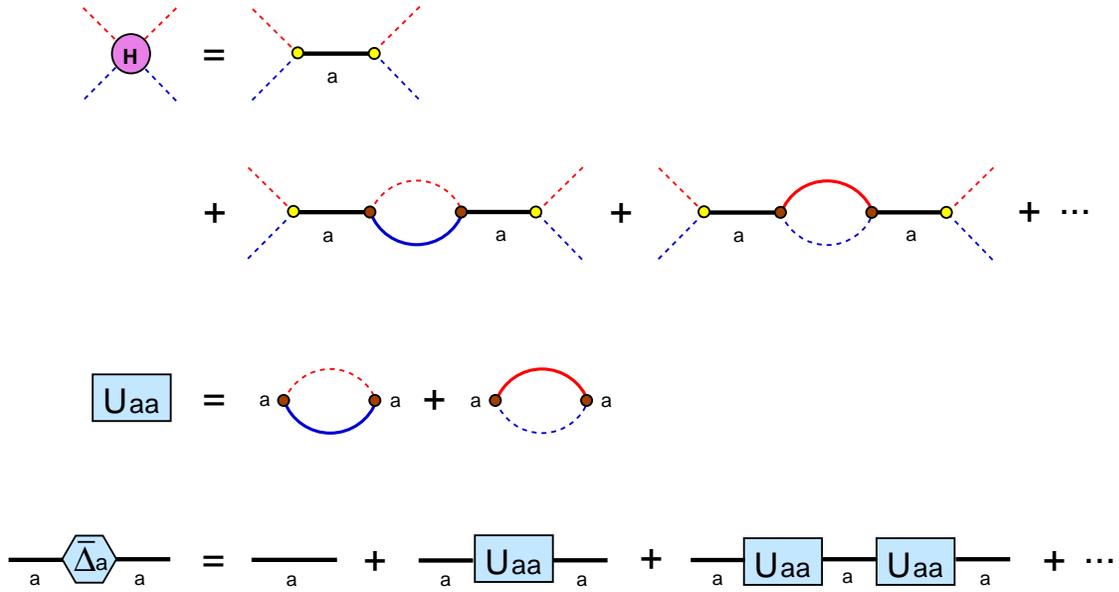}
\caption{Top: heavy kernel for a single resonance $R_a$;
continuous lines in the loops represent excited states $X^*$ or $Y^*$,
whereas the couplings $\l_a$ and $g_{a1}$ are
indicated respectively by open yellow and  black circles.
Center: heavy loop function $U_{aa}$.
Bottom: $R_a$ propagator dressed by heavy loops.
}
\label{F-H1}
\end{figure}

The expression for $\cK_H$, in the case of a single heavy 
resonance $R_a$, coupled to a single intermediate heavy loop
propagator $\o_1$,
is a direct generalization of eqs.(\ref{a8}-\ref{a10}) and
given by 
\bea
\cK_H= \cK_a  &\!=\!& 
\lc \l_a\, |\bq|\,\rc  \lc \Db_a \rc \lc \l_a\, |\bq|\,\rc \;,
\label{b1}\\[2mm]
\Db_a &\!=\!& \lc \D_a - \D_a \,U_{aa} \, \D_a 
+ \D_a \,U_{aa}\, \D_a \,U_{aa}\, \D_a + \cdots \,\rc 
= \frac{\D_a}{1 + U_{aa}\,\D_a}  \;,
\label{b2}\\[2mm]
U_{aa} &\!=\!& \lc g_{a1}\, |\bQ_1|\,\rc \lc \o_1 \rc
\lc g_{a1}\, |\bQ_1|\,\rc \;,
\label{b3}
\eea
where $\l_a$ is  the $R_a X Y$ coupling constant, 
$g_{a1}$ is the coupling constant of $R_a$ to the heavy loop
and the variable $\bQ_1$ is three-momentum inside the heavy loop.
In case the resonance can couple to two different intermediate states,
labelled by $1$ and $2$, eq.(\ref{b3}) becomes
\bea
U_{aa} &\!=\!& \lc g_{a1}\,|\bQ_1|\rc\, \lc \o_1 \rc
\lc g_{a1}\,|\bQ_1|\rc 
+ \lc g_{a2}\,|\bQ_2|\rc \lc \o_2\rc \lc g_{a2}\,|\bQ_2|\rc 
\label{b4}
\eea
and is illustrated in Fig.\ref{F-H1},
for $X Y^*$ and $X^* Y$ intermediate states.

\begin{figure}[h]
\hspace*{-120mm}
\includegraphics[width=.26\columnwidth,angle=0]{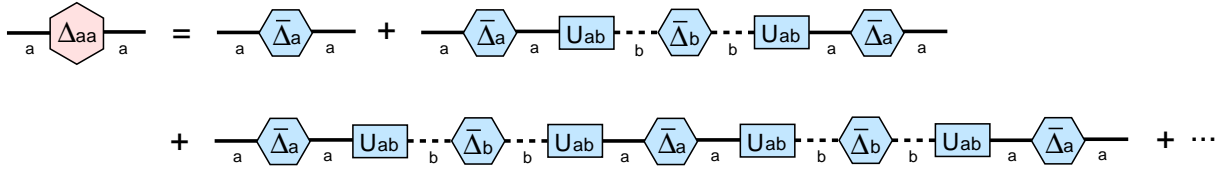}
\caption{$R_a$ propagator dressed by heavy loops and 
including forth and back oscillations into $R_b$.}
\label{F-H2}
\end{figure}

The inclusion of a second resonance $R_b$ into the problem 
gives rise to oscillations,
due to mixed couplings of the form
\bea
U_{ab} &\!=\!& \lc g_{a1}\,|\bQ_1|\rc\, \lc \o_1 \rc
\lc g_{b1}\,|\bQ_1|\rc 
+ \lc g_{a2}\,|\bQ_2|\rc \lc \o_2\rc \lc g_{b2}\,|\bQ_2|\rc \;.
\label{b5}
\eea
The full propagator describing a resonance $R_a$, which oscillates
and still ends up as a resonance $R_a$,
is represented in Fig.\ref{F-H2} and corresponds to the series
\bea 
\bD_{aa} &\!=\!& \Db_a + \Db_a \, U_{ab} \, \Db_b \, U_{ba} \, \Db_a
+ \Db_a \, U_{ab} \, \Db_b \, U_{ba} \, \Db_a \, 
U_{ab} \, \Db_b \, U_{ba} \, \Db_a + \cdots
\nn\\[2mm]
&\!=\!& \frac{\Db_a}{1- U_{ab}^2 \, \Db_a \, \Db_b} \;.
\label{b6}
\eea   
%

The propagator for a state $R_a$, which oscillates and
becomes a state $R_b$, is obtained by multiplying
eq.(\ref{b6}) by $-U_{ab}\,\Db_b$ and reads
\bea 
\bD_{ab} &\!=\!& - \Db_a U_{ab} \, \Db_b 
- \Db_a \, U_{ab} \, \Db_b \, U_{ba} \, \Db_a \, U_{ab} \, \Db_b 
- \Db_a \, U_{ab} \, \Db_b \, U_{ba} \, \Db_a \, 
U_{ab} \, \Db_b \, U_{ba} \, \Db_a \, U_{ab} \, \Db_b + \cdots
\nn\\[2mm]
&\!=\!& - \bD_{aa} \, U_{ab}\,\Db_b
= -\, \frac{\Db_a\,U_{ab} \, \Db_b}{1- U_{ba}^2 \, \Db_a \, \Db_b} \;.
\label{b7}
\eea   

\begin{figure}[h]
\hspace*{-80mm}
\includegraphics[width=.35\columnwidth,angle=0]{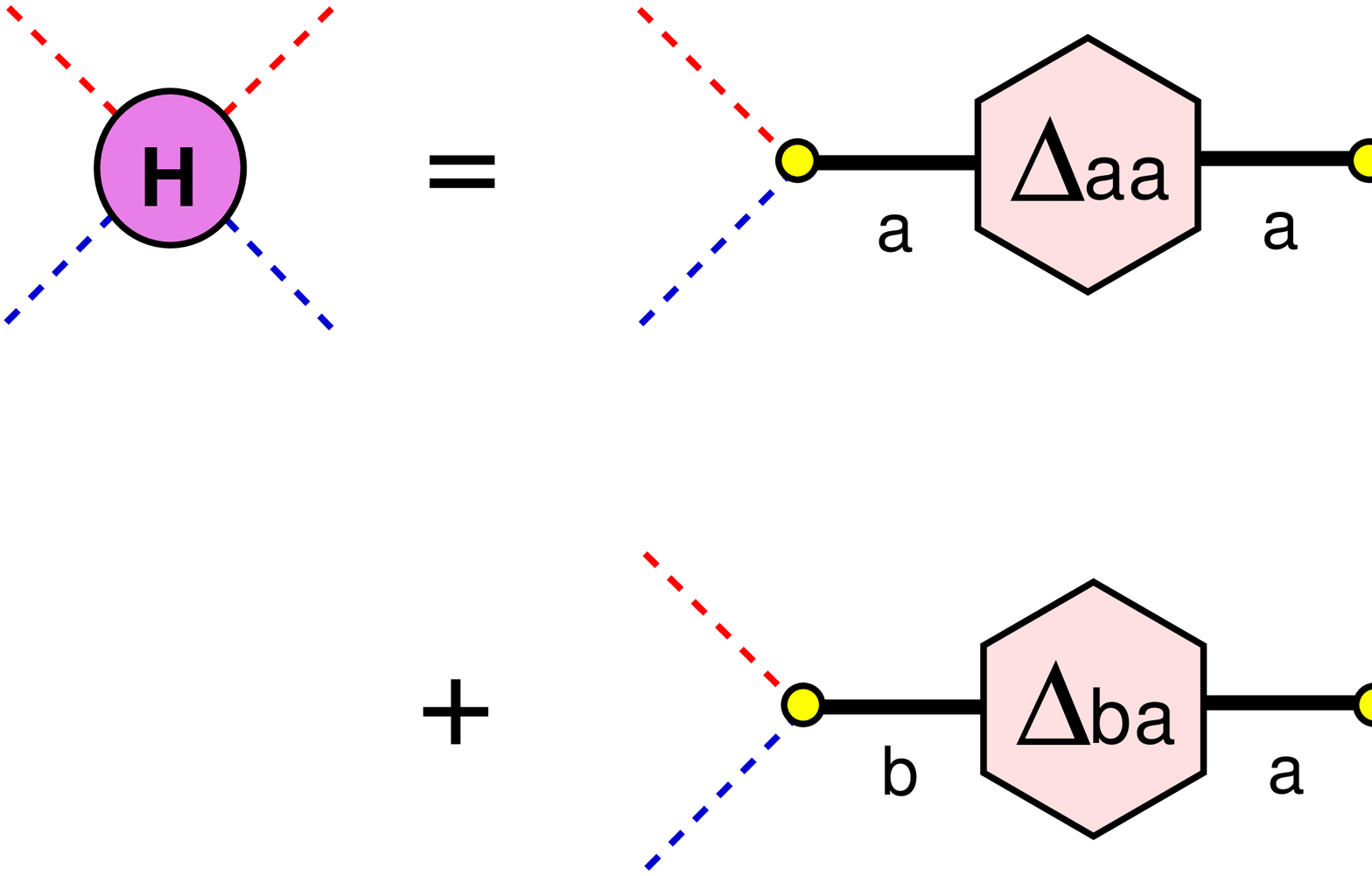}
\caption{Full heavy kernel $\cK_H$.}
\label{F-H3}
\end{figure}

The heavy component of the kernel is then given by the
diagrams of Fig.\ref{F-H3}, which correspond to
\bea
\cK_H = |\bq|^2\, \frac{\l_a^2 \, \Db_a + \l_b^2 \, \Db_b
- 2\,\l_a\,\l_b \, \Db_a \,U_{ab} \, \Db_b }
{1- U_{ba}^2 \, \Db_a \, \Db_b} \;.
\label{b8}
\eea

This result is comprehensive but may be 
too involved to be used as a practical fit function.
We therefore consider two kinds of simplifications,
among many other possible.
The first one consists in neglecting oscillations
between $R_a$ and $R_b$, by making $g_{a2}=g_{b1}=0$.
This yields $U_{ab}\rar 0$ and
\bea
\cK_H \simeq |\bq|^2 \,\lb  \l_a^2 \, \Db_a + \l_b^2 \, \Db_b \rb \;,
\label{b9}
\eea
which represents a simple sum of independent Breit-Wigner functions.
Owing to its simplicity, this choice is often made in data
analyses and sometimes refered to as {\it isobaric model}.  

In the second approximation, the coupling between $R_a$ and $R_b$
is kept, but one assumes $g_{a2}=g_{b2}=0\,$, $\l_a = \xi \,g_{a1}$ and
$\l_b = \xi \,g_{b1}$.
This gives rise to 
\bea
\cK_H \simeq  \xi^2 \, |\bq|^2\, 
\frac{g_{a1}^2 \,\D_a + g_{b1}^2\, \D_b}
{1+ \o_1 \, \bQ_1^2 \, [g_{a1}^2\,\D_a + g_{b1}^2\,\D_b]} \;.
\label{b10}
\eea
The signature of this pattern of composition is that the complex
function $\cK_H$ has to vanish at 
$s=(g_{a1}^2 m_b^2 + g_{b1}^2 m_a^2)/(g_{a1}^2 \sp g_{b1}^2)$.

The use of the so called K-matrix approximation 
yields a complementary useful simplification for $\cK_H$.
It amounts to assuming that intermediate particles are on-shell
and neglecting real parts of  heavy propagators.
This corresponds to employing
\bea
\o_i \simeq i\,[\o_1]_I = -i\, \frac{\bQ_i}{8\p\sqrt{s}}\;.
\label{b11}
\eea

\begin{figure}[h]
\hspace*{-40mm}
\includegraphics[width=0.45\columnwidth,angle=0]{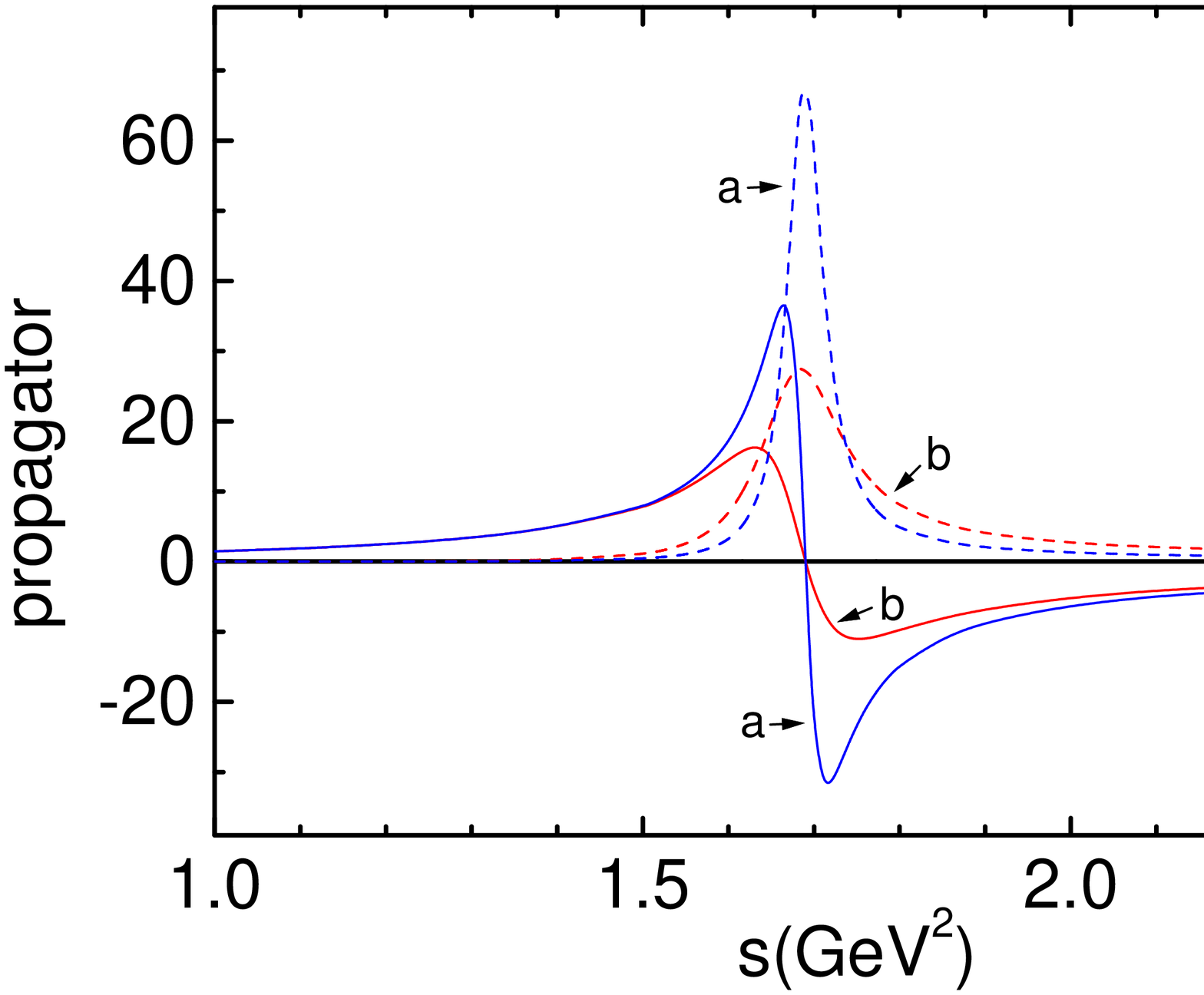}
\caption{
Real (continuous line) and imaginary (dashed line) components 
of the single resonance propagator given by eq.(\ref{b2})
for two different values (a,b) of the coupling constant $g_{a1}$;
scales are arbitrary.}
\label{F-BWS}
\vspace{5mm}
\hspace*{-40mm}
\includegraphics[width=0.45\columnwidth,angle=0]{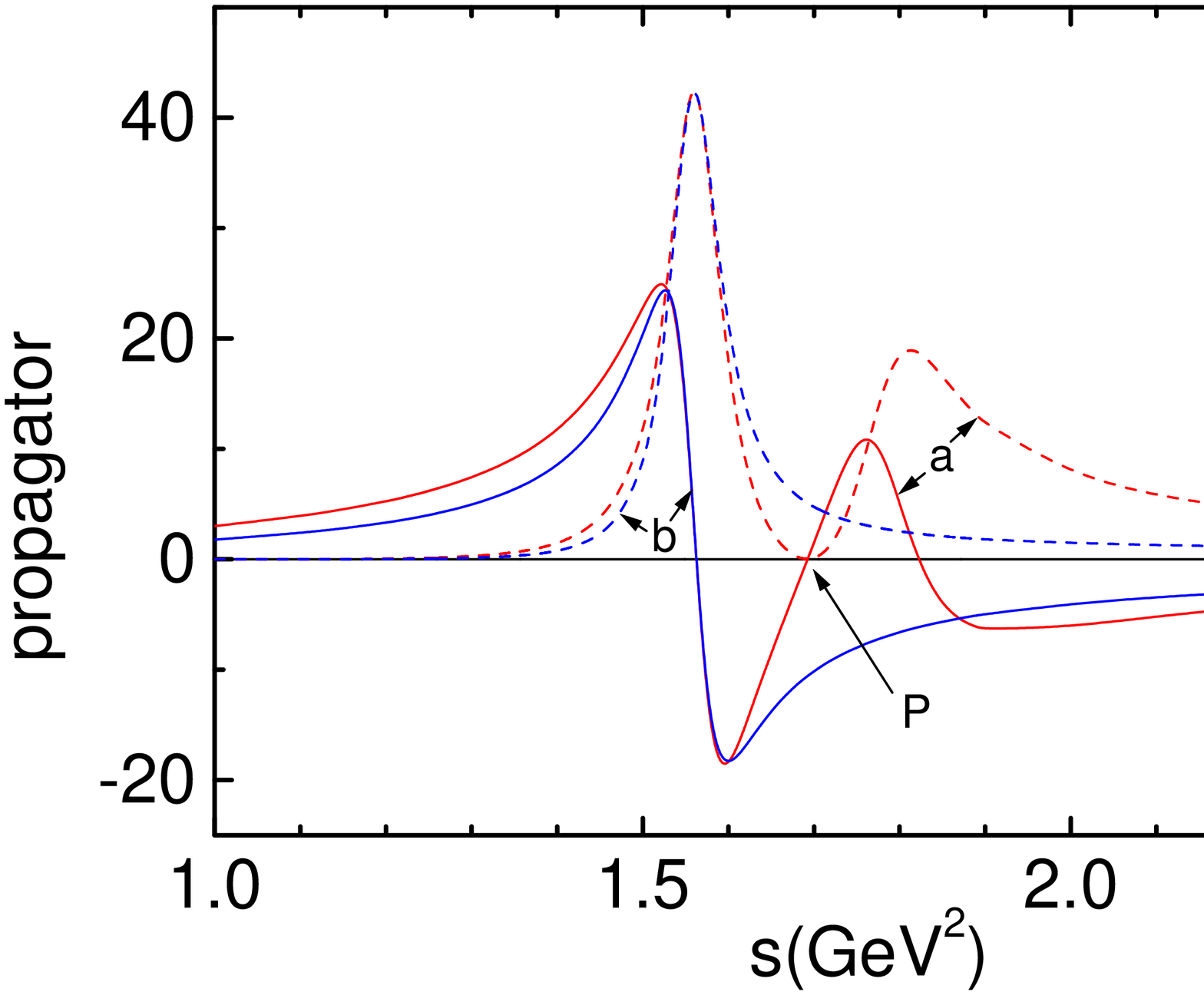}
\caption{
Real (continuous line) and imaginary (dashed line) components 
of the two-resonance propagator (red curves a), given by eq.(\ref{b10}),
compared with the case of a single resonance propagator (blue curves b).}
\label{F-BWD}
\end{figure}

We conclude by discussing the main qualitative features 
of results presented. 
In fig.\ref{F-BWS} we show the behaviour of the single resonance 
propagator, given by eq.(\ref{b2}), for two different values of the
coupling constant $g_{a1}$.
The zero of the real parts occur at $s=m_a^2$, where 
the imaginary parts reach their maxima.   
It is worth noting that the imaginary components are not symmetric
around this point.

Consequences from the approximation associated with eq.(\ref{b10})
can be seen in the curves of fig.\ref{F-BWD} where,
for the sake of comparison, the case of a single resonance 
is also shown in curves b. 
The whole two-resonance propagator vanishes at point P, 
where the real component crosses the axis and the 
imaginary component just touches it, 
since it cannot be negative.
This kind of pattern for the imaginary part might, in principle,
be related with the behaviour of $\cK_{K\p}^{exp}$, 
at point $D$ of fig.\ref{F-Kexp}.
It is interesting to compare the imaginary parts of curves $a$ and $b$
around the first peak for, even there, 
the influence of the second resonance can be seen.

\section{background $K\p$ amplitude}
\label{BG}

We describe below a possible model suited to the case of 
elastic $K\p$ scattering, inspired in chiral perturbation 
theory\cite{GL,EGPR}.
The regular part of the background kernel, which is real,
is assumed to be given by 
\bea
\cK_{{BG}_L} = \cK_L^C + \cK_L^S + \cK_L^V \;,
\label{c1}
\eea  
where $\cK_L^C$ is a leading order contact interaction,
whereas $\cK_L^S$ and $\cK_L^V$ are corrections due to 
$t$- and $u$-channel exchanges 
of scalar and vector resonances.
Using the notation of ref.\cite{EGPR}, the $P$-wave components
of these contributions are given by 
\bea
\cK_L^C &\!=\!& \frac{1}{2F^2} \; \bq^2 \;,
\label{c2}\\[2mm]
\cK_L^S  &\!=\!& 
-\, \frac{4}{F^4}
\lc \frac{ 2 \ct_d^2 \,\bq^2}{3} 
+ \lb \ct_d \, m_0^2 - 2(\ct_d \sm \ct_m)M_\p^2 \rb
\lb \ct_d m_0^2 - 2(\ct_d \sm \ct_m)M_K^2 \rb \,I_P^t(\bq^2; m_0^2) \rc
\nn\\[2mm]
&\!+\!&   \frac{1}{3F^4}
\lc \frac{ 2 c_d^2\, \bq^2}{3}  
+ \lb c_d \, m_8^2 - 2(c_d \sm c_m)M_\p^2 \rb
\lb c_d \, m_8^2 - 2(c_d \sm c_m)M_K^2 \rb I_P^t(\bq^2 ; m_8^2) \rc
\nn\\[2mm]
&\!+\!&  \frac{1}{2F^4}\;
\lc - \frac{ 2 c_d^2\, \bq^2}{3}  
+ \lb c_d \,m_{K_0^*}^2 
\sm  (c_d \sm c_m) \,(M_\p^2 \sp M_K^2) \rb^2 \, 
I_P^u(\bq^2;m_{K_0^*}^2 ) \rc\;,
\label{c3}\\[2mm]
\cK_{L\rho}^V &\!=\!& 
- \lb \frac{G_V}{F^2} \rb^2
\lc  \frac{ 2 \, \bq^2}{3} 
+ m_\rho^2 \, \lb 2(s \sm M_\p^2 \sm M_K^2) + m_\rho^2 \rb \,
I_P^t(\bq^2;m_\rho^2) \rc \;,
\\[2mm]
\cK_{L \,K^*}^V &\!=\!& - \frac{1}{4} \, \lb \frac{G_V}{F^2} \rb^2
\lc  - \frac{ 2 \, \bq^2}{3}  
+ \lb m_{K^*}^4 \sp 2m_{K^*}^2(s \sm M_\p^2 \sm M_K^2) 
\sm (M_\p^2 \sm M_K^2)^2 \rb \, I_P^u(\bq^2; m_{K^*}^2) \rc \;,\nn\\
\label{c4}
\eea

\bea
I_P^t(\bq^2; m^2) &\!=\!& \frac{1}{2 \bq^2}\, 
\lc 1 - \lb \frac{m^2}{4 \bq^2}+ \frac{1}{2} \rb \, 
\ln \lb 1 + \frac{4\bq^2}{m^2} \rb \rc \;,
\label{c5}\\[4mm]
I_P^u(\bq^2; m^2) &\!=\!& -\,\frac{1}{2\bq^2}\,
\lc 1 + \lb \frac{m^2 \sp s\sm 2(M_\p^2 \sp M_K^2)}{4\bq^2} -\frac{1}{2} \rb
\ln \lb 1 - \frac{4\bq^2}{m^2 \sp s\sm 2(M_\p^2 \sp M_K^2)} \rb \rc \;,\nn\\
\label{c6}
\eea
where $\bq$ is the three-momentum in the center of mass,
$F$, $\ct_{d,m}$, $c_{d,m}$ and $G_V$ are 
coupling constants.

\section{numerical input}
\label{values}
The  fit on $P$-wave $K\p$ amplitude with isospin $1/2$ gives  $\chi^2/n.d.f. = 2.13$ for the phase and $\chi^2/n.d.f. =5.16$ for the modulus. All  values used in this work are expressed bellow.

\ni
$\bullet$ Masses extracted from PDG 2012\cite{PDG12}  in  GeV:
$
M_\p=0.13957\,$,\, $M_K=0.493677\,$,\, $
m_\rho=0.77549\,$,\, $m_{K_0^*}=1.4250$ . 

\ni
$\bullet$ Low-energy parameters  in  GeV:  
$
F=0.10272 $  \cite{JOP},
$
G_V=0.0930/\sqrt{2}$ \cite{EGPR},
$c_d=0.0320$  \cite{EGPR}, $c_m=0.0420$\cite{EGPR}, $\ct_{d,m} = c_{d,m}/\sqrt{3}$ \cite{EGPR},
$m_0=m_8=0.9830$  \cite{GL}.

\ni
$\bullet$ Parameters extracted from $K\p$ scattering data\cite{LASS}:
$m_{K^*}=0.894$ GeV , $C=0.005127$.

\ni
$\bullet$ Fitted parameters with errors in $\cK_H$ eq.(\ref{m15}):
$
m_a=17.80354 \,(0.60089)$  GeV, $g_a^2=2.80157  \,(0.51899)\times 10^6$  GeV$^{-2}$, 
$h_a = 1.18961 \, (0.00918)$  GeV, 
$\l_a^2= 4.13935  \,(0.29267) \times 10^3$  GeV$^{-2}$, 
$m_b=1.64987 \,(0.03025) $  GeV , $
m_c=2.12355 \,(0.01283) $  GeV, $
g_b^2=12.50453 \,(0.58872) $  GeV$^{-2}$, 
$g_c^2=26.80905 \,(1.3371) $  GeV$^{-2}$,
$\xi=0.22492\,(0.00848) $, $h_b=0.8 \,(0.02211) $  GeV,
where $h$ is related to the heavy mass  by 
\bea
Q = \frac{\sqrt{s}}{2}\lp 1 - \frac{h^2}{s} \rp,
\eea
 which is an approximation of eq.(\ref{a4}) when one of the masses is considerably greater than the other.
 
\ni
$\bullet$ Fitted parameters in eq.(\ref{m17}):
$\a=1.39198 \, (0.01063)$ , 
$\b=2.52707 \, (0.09216)$, 
$\g=4.27768\, (0.20357)$ .



\end{document}